\documentstyle[epsfig]{aipproc}
\begin{document}

\pagestyle{empty}
~
\vfill
\begin{center} \setlength{\unitlength}{1cm} \begin{picture}(0,0) 
\put(-0.25,-1.0){\makebox(0,0){\epsfig{file=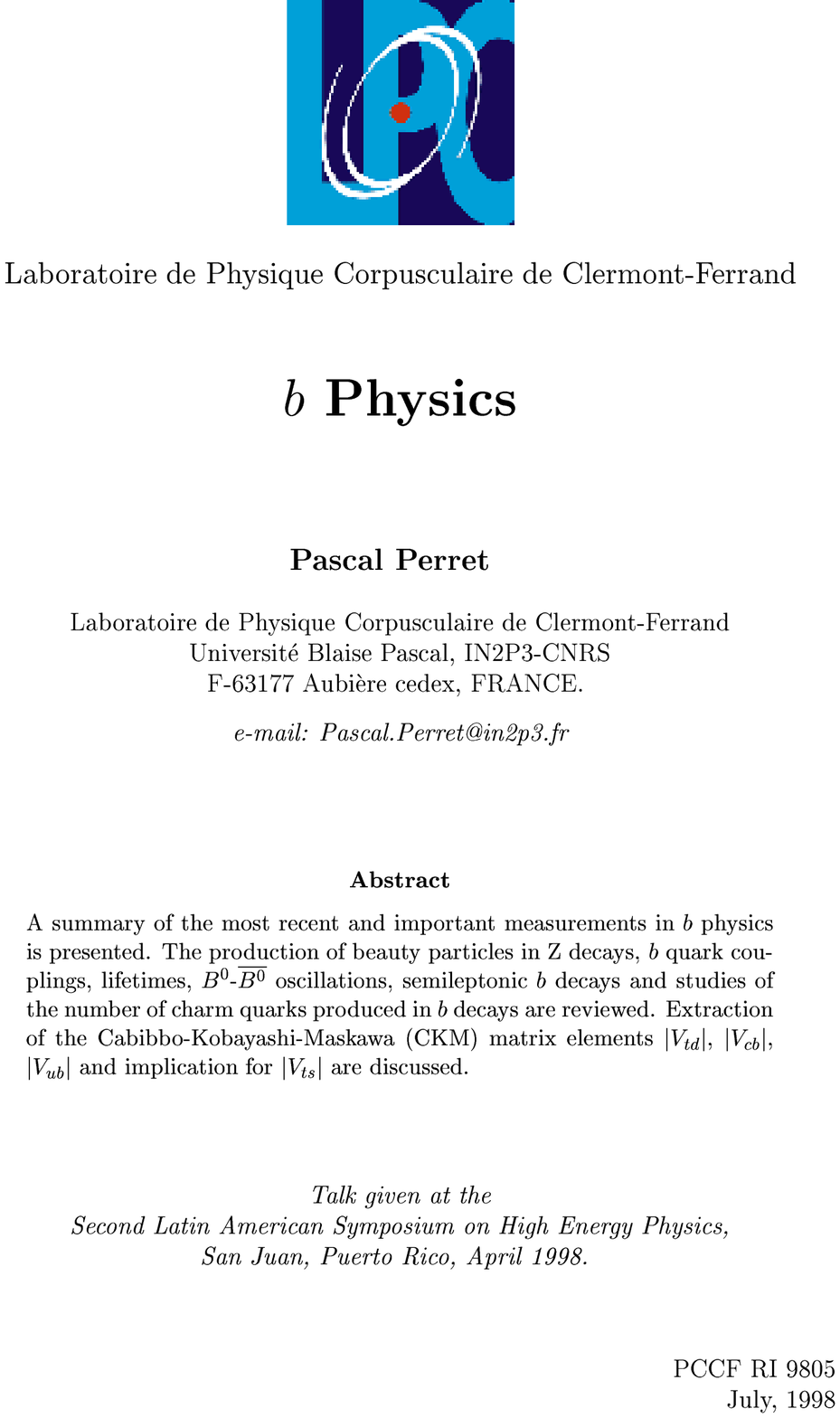,width=1.3839\textwidth}}}
\end{picture} \end{center}
\vfill
~
\newpage 

~
\newpage \pagestyle{plain} \setcounter{page}{1}

\title{$b$ Physics}
\author{Pascal Perret}
\address{Laboratoire de Physique Corpusculaire, 
Universit\'e Blaise Pascal, IN2P3-CNRS \\
F-63177 Aubi\`ere cedex, FRANCE. 
}
\maketitle

\renewcommand{\b}{\mbox{$B^0$}}
\newcommand{\bq}{\mbox{$B^0_q$}}
\newcommand{\bs}{\mbox{$B^0_s$}}
\newcommand{\bd}{\mbox{$B^0_d$}}
\newcommand{\bbar}{\mbox{$\overline{B^0}$}}
\newcommand{\bqbar}{\mbox{$\overline{B^0_q}$}}
\newcommand{\bsbar}{\mbox{$\overline{B^0_s}$}}
\newcommand{\bdbar}{\mbox{$\overline{B^0_d}$}}
\newcommand{\dmq}{\mbox{$\Delta m_q$}}
\newcommand{\dms}{\mbox{$\Delta m_s$}}
\newcommand{\dmd}{\mbox{$\Delta m_d$}}
\newcommand{\dmdw}{\mbox{$\Delta m^{\rm world}_{\rm d}$}}
\newcommand{\chid}{\mbox{$\chi_{\rm d}$}}
\newcommand{\chis}{\mbox{$\chi_{\rm s}$}}
\newcommand{\chidw}{\mbox{$\chi^{\rm world}_{\rm d}$}}

\begin{abstract}
A summary of the most recent and important measurements in $b$ physics
is presented.
The production of beauty particles in Z decays, $b$ quark couplings,
lifetimes, \b-\bbar\ oscillations, 
semileptonic $b$ decays 
and studies of the number of charm quarks produced in $b$ decays are reviewed.
Extraction of the Cabibbo-Kobayashi-Maskawa (CKM) matrix elements $|V_{td}|$,
 $|V_{cb}|$, $|V_{ub}|$ and implication for $|V_{ts}|$ are discussed.
\end{abstract}

\section{Introduction}
The heavy mass of the $b$ quark, around 5 GeV, so much greater
than the strong interaction scale $\Lambda_{QCD} \sim 0.2$ GeV and 
the fact that it belongs to the same isospin doublet than the top quark,
confere 
a special role to the $b$ quark studies.
 Furthermore the top quark is too heavy to build
hadrons and thus $b$ hadrons are the heaviest. 
In that respect,
$b$ physics is a broad
subject and 
one of the major areas of investigation of present
experiments at CESR, LEP, SLC and Tevatron colliders.
Experimentally $b$ hadrons are easier to observe or to disentagle from
other sources because tracks issued from their decays have higher
transverse momentum and momentum due to the high mass and the 
hard fragmentation 
of the $b$ hadrons. The lifetime of $b$ hadrons ($\sim $ 1 ps) is
relatively long and the subsequent presence of secondary vertices in detector
can be used as a tag; for instance the mean decay length is 3 mm at LEP.
They have also  sizeable semileptonic branching ratios which allow to
sign $b$ events with the presence of leptons, cleanly identified,
in decay products.
Specific theoretical framework can be used for the description of the
properties of $b$ hadrons such as Heavy Quark Effective Theory (HQET)
where a $b$ hadron is considered like a hydrogen atom, Heavy Quark
Symmetry (HQS) in the limit $m_b \to  \infty$ or Heavy Quark Expansion
(HQE) which allows expansions in $1/m_b$.

The main issues in $b$ physics are 
to provide
precision tests in the electroweak sector of the Standard Model (SM),
to study the decay dynamics, especially the effect of strong interactions
 on the underlying quark decay, 
to understand the origin of CP violation and to measure the magnitude of the 
CKM matrix elements  $|V_{cb}|$, $|V_{ub}|$, $|V_{td}|$ and $|V_{ts}|$, 
and to observe rare processes which can probe physics beyond the SM.
A selection of subjects is reviewed and, due to space limitations, only
a short summary is given. More extensive recent summaries can be 
found in \cite{schneid,drell,rich,feindt}. Production in Z decays,
lifetimes, \b-\bbar\ oscillations and decays are mainly discussed in the
following. Most of the given averages are provided by LEP working
groups~\cite{LEPEWWG,LEPBLIF,LEPBOSC}.

In the field of beauty hadron spectroscopy, 
the main result is the observation and measurement of the $B_c$ meson,
by CDF at Tevatron, with a mass of $6.40 \pm 0.39  \pm 0.13$ Gev/c$^2$
and a lifetime of $0.46^{+0.18}_{-0.16} \pm 0.05$ ps. More details can be
found in the presentation of J. Troconiz, where recent results
from Tevatron are covered.

\section{$b$ production in Z decays}
The relative ease with which $b$ quarks can be separated from other quark
flavours and the availability of large $Z^0$ event samples allow precision
tests of the Standard Model to be carried out using ${\rm Z} \to b \bar b$
decays at $e^+e^-$ colliders. By the end of the LEP I phase (1989-1995),
each of the four LEP experiments had recorded approximately 3.8 10$^6$
${\rm Z} \to q \bar q$, including nearly 0.8 10$^6$ ${\rm Z} \to b \bar b$
decays, while by the end of 1997 SLD at SLC had recorded approximately 
0.3 10$^6$ ${\rm Z} \to q \bar q$ with polarized beams.

{\boldmath  $R_b$}:
Due to the large mass of the $b$ quark and the fact that it belongs to the
same isospin doublet than the top quark, the Z-$b \bar b$
coupling is one of the most interesting windows in the search for new physics.
The partial width ratio
$ R_b = \frac{\Gamma({\rm Z} \to b \bar b)}{\Gamma({\rm Z} \to q \bar q)} $
is sensitive to $m_{\rm top}$ via vertex corrections,
while the
corrections in $\alpha_s$ and $m_H$ are suppressed in a first approximation.
A precision measurement of $R_b$ would therefore represent a unique test of 
the Standard Model and would provide significant constraints on possible
new physics such as additional Higgs bosons or supersymmetry.

The first $R_b$ measurements were done using 
leptons~\cite{alrbl,delrbl,l3rbl,oprbl}. In inclusive charged
lepton analyses, the preferred approach is to fit a two-dimensional
distribution (p,p$_{\perp}$) for single lepton and dilepton events together,
and extract simulatneously $R_b$ with other parameters as for instance
$R_c$ or ${\cal B}(b \to \ell)$. Relatively large systematic errors 
remain due mainly to uncertainties in the modelling of the semileptonic decay
of the $b$ quark; these measurements were not precise enough to perform a
stringent test. Then the informations from silicon vertex detector were used 
and thanks to the large statistics available, ``double tagging'' techniques,
were applied. The double tagging technique exploits the fact that the $b$ 
and $\bar b$ quarks are typically produced back to back, in separate
hemispheres as defined by the thrust axis. A $b$ quark tag is applied
separately to each hemisphere in a sample of hadronic events and the total
number of single and double-tagged events are measured.
Assuming backgrounds from charm and light quark events
and the correlations between the hemispheres from Monte Carlo, 
as well as $R_c$ 
from its SM value, both $R_b$ and the $b$ tagging efficiency can be extracted
from data. In 1995 the world average $R_b$ showed a discrepancy with more
than 3$\sigma$ (dominated by systematics) from the SM, and it was shown that 
charm systematics were a worry (exponential charm lifetime tail ($D^+$) is 
difficult to cut away) and so a $b$ purity of 94 \% was not enough controlled,
 as well as
the understanding of the correlations. A new round of analyses has been
performed~\cite{alrb,delrb,l3rb1,sldrb}. 
An increased purity is achieved by exploiting $b/c$ hadrons masses
and kinematical differences, by including for instance 
the invariant mass of the significant tracks. 
The primary vertex was initially measured with all tracks of the event,
including a correlation between hemispheres.
Primary vertices were then reconstructed, one per hemisphere.
Finally lifetime tag was used in conjunction with other tags; variables
are combined in multivariate analyses (neural network for instance). 
With this kind of analyses, a purity greater than 98 \% with an efficiency
greater than 30 \% has been achieved by DELPHI~\cite{delrb}.
Measurements are summarised in table \ref{tabrb}. The combined 
result~\cite{LEPEWWG},
$R^0_b$ =  0.21732 $\pm$ 0.00087, 
corresponds to a precision $\frac{\Delta R_b}{R_b} = 0.4 \% $.
The statistical and systematic errors are comparable, the latter receiving
contributions mainly from uncertainties in gluon splitting $ g \to b \bar b$
and $g \to c \bar c$, from the tracking resolution of the detector,
and from hemisphere correlations.

\begin{table}
\begin{center}
\caption{Summary of $R_b$ results.}
\begin{tabular}{|lcr|l|}
\hline
ALEPH~\cite{alrb}  & mult & 1992-95 & 0.2159 $\pm$ 0.0009 $\pm$ 0.0011 \\
DELPHI~\cite{delrb} & mult & 1994-95 & 0.2166 $\pm$ 0.0008 $\pm$ 0.0009 \\
L3~\cite{l3rb1}     & mult & 1994-95 & 0.2179 $\pm$ 0.0015 $\pm$ 0.0026 \\
L3~\cite{l3rb2}     & shape & 1991 & 0.2223 $\pm$ 0.0030 $\pm$ 0.0064 \\
OPAL~\cite{oprb}   & mult & 1992-94 & 0.2178 $\pm$ 0.0014 $\pm$ 0.0017 \\
SLD~\cite{sldrb}    & vtx mass & 1993-97 & 0.2158 $\pm$ 0.0017 $\pm$ 0.0014 \\
\hline
LEP~\cite{alrbl,delrbl,l3rbl,oprbl}    & leptons &  & 0.2227 $\pm$ 0.0020 $\pm$ 0.0025 \\
\hline
\hline
LEP + SLC &\multicolumn{2}{c|}{corrected for $\gamma$ exchange}  
& 0.21732 $\pm$ 0.00087 \\
\hline
\end{tabular}
\label{tabrb}
\end{center}
\end{table}

{\bf Gluon splitting:} The gluon splitting $g \to b \bar b$ is an important
ingredient in the $R_b$ measurement, constituting the largest single
systematic uncertainty. New methods have been developed to measure this
parameter by searching for $b$-tagged jets in 4 jet events. The 2 $b$-tagged 
jets have to form a small angle and the initial quarks are required to be
in opposite hemispheres. DELPHI has measured $g \to b \bar b$ = (0.21
$\pm$ 0.11 $\pm$ 0.09)\% \cite{Dgbb}
and ALEPH (0.26 $\pm$ 0.04 $\pm$ 0.09)\% \cite{Agbb}, the average
being (0.24 $\pm$ 0.09)\% .

{\bf {\boldmath $b$} Asymmetry:} The other electroweak quantity of interest 
is the forward
backward charge asymmetry  $A^b_{FB}$, obtained from measurements of the
angular distribution $\frac{d \sigma}{d \cos\theta} \propto 1 + \cos^2\theta + 
\frac{3}{8}A^b_{FB} \cos \theta$; where $\theta$ is the angle of the outgoing 
$b$ quark with respect to the initial $e^-$ direction. The asymmetry  
$A^b_{FB}$ arises from differences in the coupling strengths of the Z to
left- and right-handed fermions, and is one of the most sensitive quantities 
to the effective electroweak mixing angle $\sin^2 \theta_{\rm eff}^{\rm lept}$
= $1/4(1-g_{V\ell}/g_{A\ell})$~\cite{LEPEWWG}.
To measure $A^b_{FB}$, one needs to select a
$b$ sample, to define the $b$ quark direction 
(usually approximated by the thrust
axis), and to estimate the electric charge of the quark to 
assign the $b$ quark to the forward or backward direction.
Leptons are good candidates, high p and p$_{\perp}$ leptons come mainly
from $b$ quarks and their electric charge allows to identify the $b$ quark
hemisphere~\cite{alas,delas,l3as,opas}. 
Tag of Z $\to b \bar b$ using a lifetime/mass tag and using 
a momentum weighted track charge in each hemisphere to flag the $b$ quark 
is also performed~\cite{alasj,delas,l3asj,opasj}.
Both approaches are still statistically limited and achieve a similar overall
precision. 
$D^{* \pm}$ or $K^{\pm}$ tag can also be used but are less 
performing~\cite{delas,opasd}.
For instance, at SLD,
$b \bar b$ events are tagged using a mass tag, while kaons from
$b \to c\to s \to K$ are used to sign the $b$ quark direction with their
charge and direction.
The main measurements are summarized in table~\ref{tabab}, and the
average of the pole asymmetry is $A^{0,b}_{FB}$ = 0.0998 $\pm$ 0.0022 \cite{LEPEWWG}.
$A^{0,b}_{FB}$ can be expressed as a measurement of the effective 
angle $\theta_{\rm eff}^{\rm lept}$:
 $\sin^2 \theta_{\rm eff}^{\rm lept}$
= 0.23213 $\pm$ 0.00039 .

\begin{table}
\begin{center}
\caption{Summary of $A^b_{FB}$ results at $\sqrt{s} 
\approx m_{\rm Z}$. }
\begin{tabular}{|lcr|l|}
\hline
ALEPH~\cite{alas}  & leptons & 1990-95 & 0.0965 $\pm$ 0.0044 $\pm$ 0.0026 \\
DELPHI~\cite{delas} & leptons & 1991-94 & 0.1075 $\pm$ 0.0077 $\pm$ 0.0031 \\
L3~\cite{l3as}     & leptons & 1990-95 & 0.0963 $\pm$ 0.0065 $\pm$ 0.0035 \\
OPAL~\cite{opas}   & leptons & 1990-95 & 0.0910 $\pm$ 0.0044 $\pm$ 0.0020 \\
\hline
ALEPH~\cite{alasj}  & jet-charge  & 1991-95 & 0.1017 $\pm$ 0.0038 $\pm$ 0.0032 \\
DELPHI~\cite{delas} & jet-charge  & 1991-94 & 0.0995 $\pm$ 0.0072 $\pm$ 0.0040 \\
L3~\cite{l3asj}     & jet-charge  & 1994    & 0.0855 $\pm$ 0.0118 $\pm$ 0.0056 \\
OPAL~\cite{opasj}   & jet-charge  & 1991-95 & 0.1004 $\pm$ 0.0052 $\pm$ 0.0046 \\
\hline
\hline
LEP + SLD   &$A^{0,b}_{FB}$ & winter 98  & 0.0998 $\pm$ 0.0022 \\
\hline
\end{tabular}
\label{tabab}
\end{center}
\end{table}

\section{Lifetimes}
In the quark spectator model, the heavy quark decays weakly without
interacting with the other light quark(s). As a result, all the hadrons
containing a $b$ quark should have the same lifetime.
As in the case of the charm hadrons, non-spectator effects, such as
final state interference, W exchange, weak annihilation and helicity
suppression lead to significant differences in the lifetimes of beauty hadrons.
In heavy quark expansion (HQE) theory, a theoretical approach based on QCD and
where the decay rates of a beauty hadron are expressed as an expansion in
powers of 1/$m_b$, the lifetime difference of baryons and mesons depends
on terms of the order of $1/m_b^2$ and higher, while the lifetime of the
different B mesons depend on terms $1/m_b^3$~\cite{HQE}. 
The following hierarchy among the various species 
$\tau_{\Lambda_b} < \tau_{B^0_d} \simeq  \tau_{B^0_s} <  \tau_{B^{+}}$
is expected~\cite{lifpred}, 
but it seems that corrections in ``${\cal O}(1/m_b^3)$'' could
be large in the ratio $\tau(B^+)/\tau(B^0)$ without model 
assumptions~\cite{Bigi_lifetimes,Neubert_lifetimes}.
The experimental determination of the magnitude of these differences
is needed.

To measure the proper lifetime of a B hadron, it is necessary to 
determine its decay length and its momentum. 
Several different and complementary
methods have been developed to perform such measurements. Fully reconstructed
beauty hadron final states are the cleanest way. These 
measurements~\cite{CDF_web} benefit 
from the precise determination of the secondary vertex, and since there are
no missing particles, the momentum is well determined. Consequently
these measurements have little dependence on simulation. However this
technique is limited at LEP/SLC due to the available statistics.
Larger samples are obtained by using the presence of a high
momentum lepton to select  semileptonic $b$ decays, and by fully or partially
reconstructing a charm hadron of the appropriate charge in the same 
jet~\cite{CDF_web,OPAL_tau_Bs_Lamb_EPS,DELPHI_taus_dms_EPS,DELPHI_inclDstl,L3_tau_Bzero_EPS,ALEPH_tau_Lamb,DELPHI_tau_Lamb,OPAL_tau_Dshad}.
The vertex resolution is still good due to the lepton, but the missing 
products degrade the momentum resolution. These methods suffer also 
from higher background due to fake contaminations and the ``pollution''
of B$^+$ and B$^0$ cross-contamination for instance.
Another approach is based on pure topological vertexing. $b$ decay
vertices are reconstructed inclusively and the $b$ hadron charge is
determined from the total charge of the tracks associated with its 
vertex~\cite{SLD_topol,SLD_Warsaw}.
This method gives the highest statistics at the expense of a reduced purity
and a greater sensitivity to the modelling simulation.
Some measurements of the $b$ lifetime over all hadron species are based
on the impact parameters of tracks from $b$ decays, generally leptons.
The knowledge of the $b$ fragmentation and of the semileptonic decay models
systematically limits the accuracy of these measurements.

There are many lifetime measurements, their average~\cite{LEPBLIF} is given in 
table \ref{tablif}.
Lifetime ratios are known experimentally close to 5\% and the lifetime
hierarchy among beauty hadrons is predicted correctly. No significant
differences between the three B mesons lifetimes are observed and
they are in good agreement  with HQE predictions. However the ratio
 $\tau(b-{\rm baryon})/\tau(B^0)$ is significantly different from 
unity~\cite{ALEPH_tau_Lamb,DELPHI_tau_Lamb,OPAL_tau_Bs_Lamb_EPS,l3mult}
and smaller than usual predictions. This is correlated with a small
semileptonic $b$-baryon semileptonic branching ratio (see below),
and is the place of an intensive work.

\begin{table}[h]
\begin{center}
\caption{World average lifetime measurements and their ratio. 
Predictions
are given in the last column.}
\begin{tabular}{|lc||lc|c|}
\hline
$\tau(B^+)$ & 1.67 $\pm$ 0.04 ps & $\tau(B^+)/\tau(B^0)$ & 1.07 $\pm$ 0.04 
& 1.0 - 1.1 \\
$\tau(B^0)$ & 1.57 $\pm$ 0.04 ps & & & \\
$\tau(B_s)$ & 1.48 $\pm$ 0.06 ps & $\tau(B_s)/\tau(B^0)$ & 0.95 $\pm$ 0.05 
& 0.99 - 1.01 \\
$\tau(\Lambda_b)$ & 1.23 $\pm$ 0.08 ps & $\tau(\Lambda_b)/\tau(B^0)$ 
& 0.78 $\pm$ 0.06 & 0.9 - 1.0 \\
$\tau(b-{\rm baryon})$  &  1.22 $\pm$ 0.05 ps & $\tau(b-{\rm baryon})/\tau(B^0)$ 
& 0.78 $\pm$ 0.04 & 0.9 - 1.0 \\
\hline
\hline
$\tau(b)$ & 1.554 $\pm$ 0.013 ps & & & \\
\hline
\end{tabular}
\label{tablif}
\end{center}
\end{table}


\section{\b-\bbar\ oscillations}
In the Standard Model, particle-anti-particle oscillations take place
via a second order weak interaction process - box diagram - with a loop
of W bosons and up-type quarks, which are dominated by top quark 
exchange in the case of neutral B mesons. The oscillation frequency
depends on the mass difference \dmq\ between the mass eigenstates.
Time integrated measurements are performed, they are typically based
on counting same-sign and opposite-sign lepton pairs. At LEP both 
neutral B meson species are produced with a rate $f_{B^0_d}$ and $f_{B^0_s}$
(see below), and the LEP average is  \mbox{$\overline \chi$} = 
$f_{B^0_d} \chi_d + f_{B^0_s} \chi_s$ =  0.1214 $\pm$ 0.0043~\cite{LEPEWWG}, 
while CLEO and ARGUS,
at the $\Upsilon(4s)$ where only \bd\ mesons are produced, measured
$\chi_d^{\Upsilon(4s)}$ = 0.156 $\pm$ 0.024~\cite{PDG}.

To measure the time dependence of the mixing, one needs to know the
$b$ flavour at production time and at decay time to define whether 
a mixing occured or not, as well as the B decay length and energy to 
reconstruct the proper decay time. Many different methods have been
developed for this purpose. The final state tag is given by the charge of
the decay products (lepton,  $D^{*\pm}$, $D_s^\pm$ or  $K^\pm$ \cite{oscsld}).
For fully inclusive analyses based on topological vertexing, the final
state tagging techniques include jet charge~\cite{oscjc} and charge dipole 
methods~\cite{oscsld}.
For the initial flavour state, we can distinguish tags which exploit
the B hadron decay in the opposite hemisphere using the charge of a lepton
or a kaon, and those which exploit informations of the B candidate itself.
These later one use the charge of a track from the primary vertex 
which is correlated 
with the production state of the B if that track is a decay product of a 
$B^{**}$ state or if it is the first particle in the fragmentation 
chain~\cite{CDF_dmd,ALEPH_dms_LPEPS}.
The jet charge techniques work on both sides. 
At SLC, the beam polarization produces a sizeable forward-backward 
asymmetry in the ${\rm Z} \to b \bar b$ decays and provides another
very interesting and effective initial state tag, based on the polar angle 
of the B candidate~\cite{oscsld}.

\begin{figure}[htbp]
\begin{center}
\epsfig{file=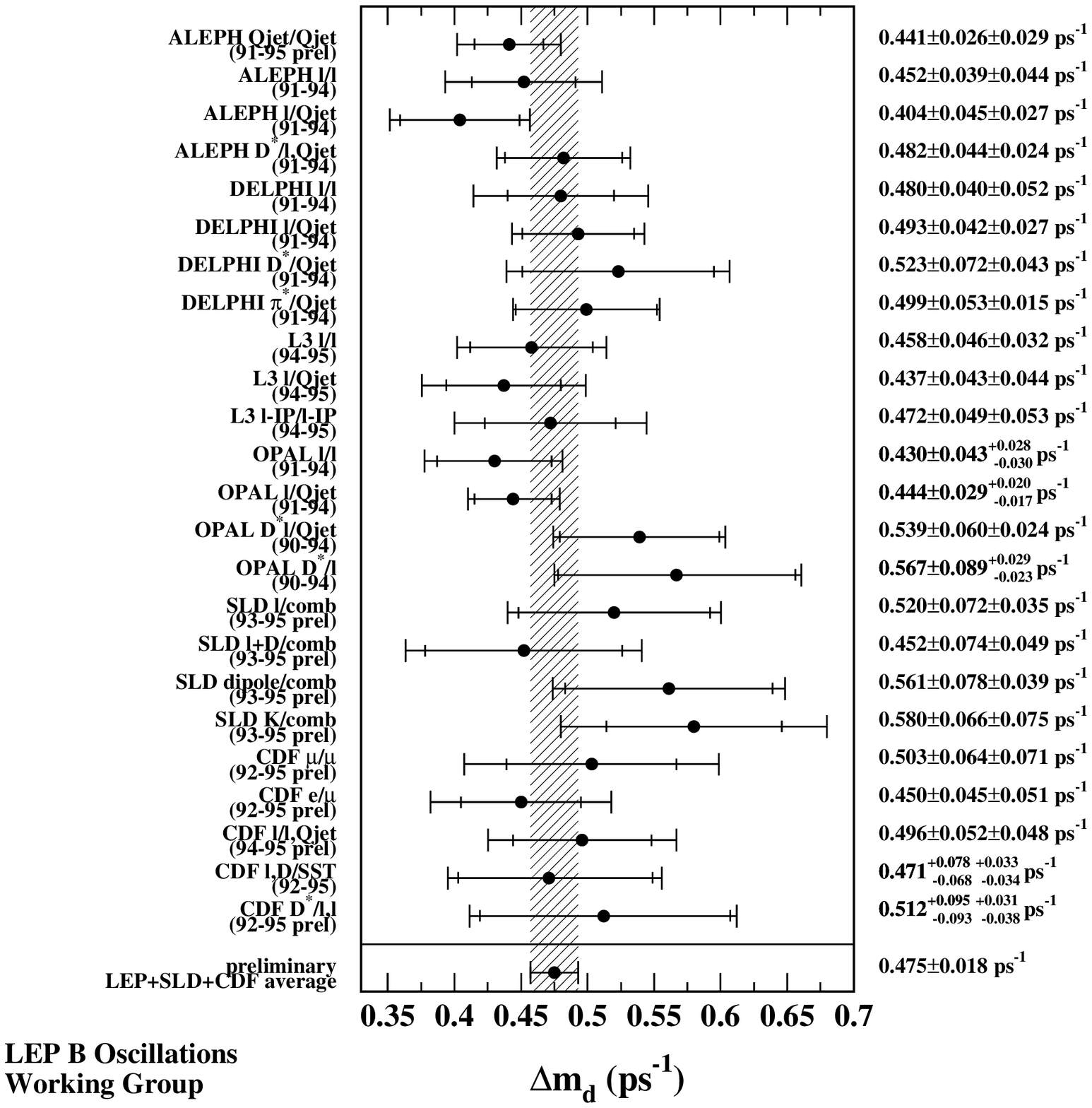,width=1.2\textwidth
,height=\textheight} 
\caption{Measurements of \dmd.}
\label{figdmd1}
\end{center}
\end{figure}

A lot of different analyses have been performed to measure 
\dmd \cite{CDF_web,oscsld,oscjc,CDF_dmd,OPAL_dmd_dms,ALEPH_dmd,DELPHI_dmd,L3_dmd,OPAL_dmd}.
An overview is shown in figure~\ref{figdmd1}.
Averaging all direct 
\dmd\ measurements from LEP, SLD and CDF,
yields $0.475 \pm 0.018{\rm ps^{-1}}$~\cite{LEPBOSC}.
The systematic uncertainties are not negligible; 
they are often dominated by the sample composition, mistag probability,
or $b$ hadron lifetime contributions.
Including CLEO and ARGUS measurements of \chid~\cite{PDG}
give the world averages~\cite{LEPBOSC}:
$\dmdw = 0.466 \pm 0.018~{\rm ps^{-1}}$ and
$\chidw=0.174\pm 0.011$.

\begin{figure}[htbp]
\begin{center}
\epsfig{file=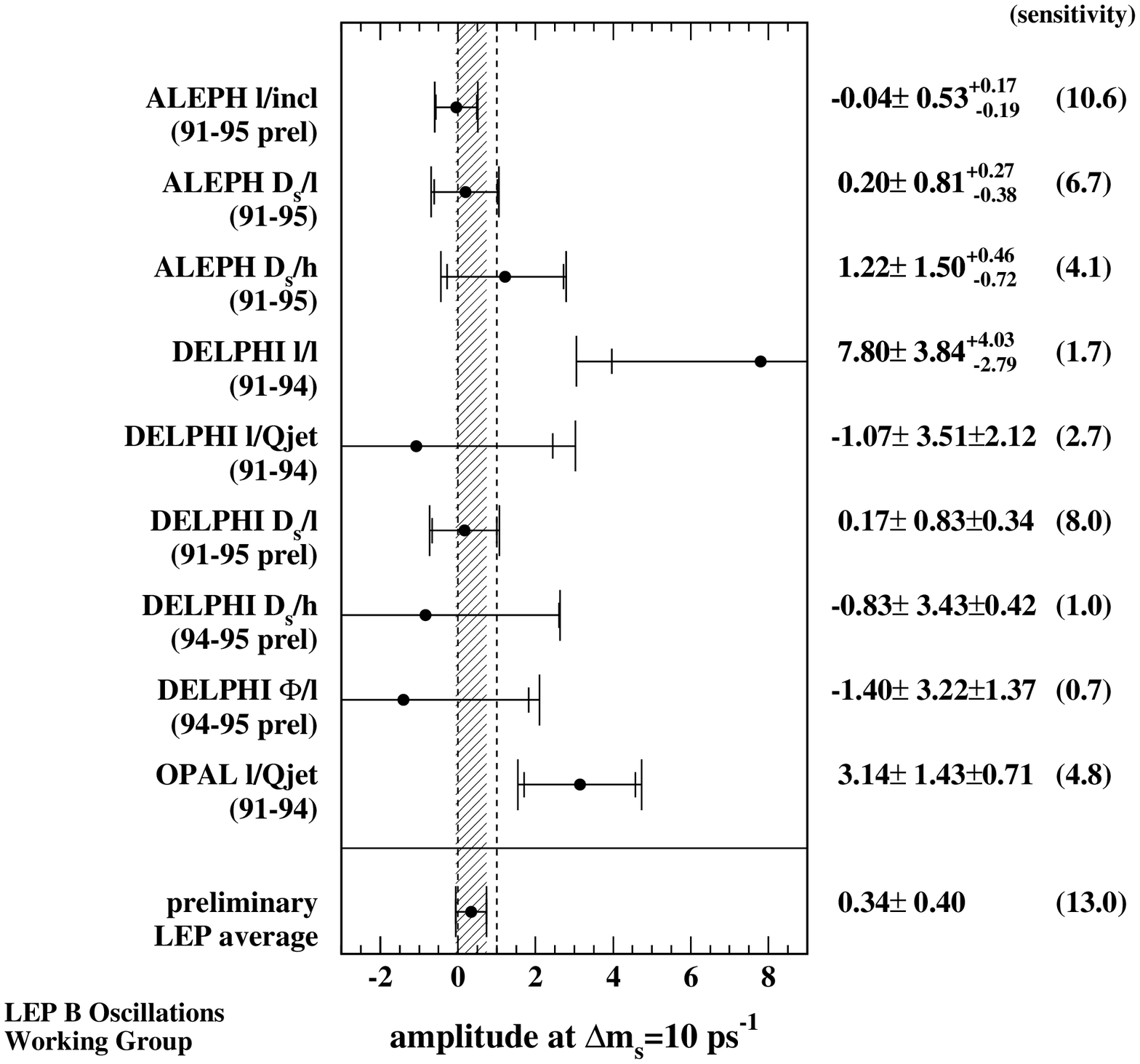,width=1.1\textwidth,height=\textheight} 
\caption{Sensitivities on \dms.}
\label{figdms}
\end{center}
\end{figure}

The $B^0_s$ oscillations have been the subject of many recent 
studies~\cite{ALEPH_dms_LPEPS,DELPHI_dms_LP,OPAL_dmd_dms}.
However, the $B^0_s$ mixing proceeds much faster than the $B^0_d$ mixing,
and the time evolution has not been resolved. Only lower limits
are derived, and an overview of the available sensitivities is
given in figure~\ref{figdms}.
The combined 95\%  Confidence Level (C.L.) limit, derived from the amplitude
method~\cite{ampli}, is \dms $> 10.2 {\rm ps^{-1}}$~\cite{LEPBOSC}.

The measurement of 
\dmd\ and \dms\ are related, in the Standard Model, to 
the CKM matrix elements $V_{td}$ and $V_{ts}$ respectively. 
From \dmd\ one gets 
$|V_{\rm td}|  =   (8.8 \pm 0.2_{\Delta m_{\rm d}}
\mp 0.2_{m_{\rm t}} {\mp^{1.4}_{1.8}}_{th}) \times 10^{-3} $, with an
uncertainty completely dominated by theoretical uncertainties.
However, many uncertainties cancel in the frequency ratio, yelding
$ |V_{\rm ts}|/|V_{\rm td}|  >   3.8$ at 95\% C.L.

The $B^0_s$ and $b$ baryon fractions can be
extracted from branching ratio measurements. 
The LEP B oscillations working group estimates~\cite{LEPBOSC}
$f_{\mbox{\scriptsize b-baryon}}$  $ = (10.6 ^{+3.7} _{-2.7}) \%$
and $f_{B^0_s}$ $ = (10.8 ^{+3.3} _{-2.9}) \%.$
$\dmdw$ and $\chidw$ can be used to 
improve our knowledge on the fractions of weakly decaying 
bottom hadron in ${\rm Z} \to b \bar b$ events.
If one assumes also $\chis = 1/2$ and
$f_{B^0}=f_{B^+} = (1-f_{B^0_s}-f_{\mbox{\scriptsize b-baryon}})/2$, another
estimate of $f_{B^0_s}$ can be extracted from \chidw, from
the inclusive integrated mixing rate $\overline{\chi}$ measured at LEP, from
the $f_{\mbox{\scriptsize b-baryon}}$  branching ratios estimate 
and from the $b$ hadron lifetimes. Combining all the informations 
yields $f_{B^0_s}=(10.3^{+1.6}_{-1.5})\%$,
$f_{\mbox{\scriptsize b-baryon}}=(10.6^{+3.7}_{-2.7})\%$ and
$f_{B^0}=f_{B^+} = (39.5^{+1.6}_{-2.0})\%$. These results, 
including \dmdw,
have been obtained by the LEP B
oscillations working group in a consistent way.
There are also new measurements of $f_{\mbox{\scriptsize b-baryon}}$ 
 $ = (10.2 \pm 0.7 \pm 2.7) \%$ from ALEPH~\cite{alfbar} and of
$f_{B^0_s}=(10.8 \pm 1.3 \pm 2.2)\%$ from DELPHI~\cite{delfbs}
(from $f_{B^0_s+B^{**}_s}$ = ($14.4 \pm 1.7 \pm 3.0$)\% and assuming
$f_{B^0_s}$/($f_{B^0_s}+f_{B^{**}_s}$) =  $0.25 \pm 0.05$) .
Including these measurements yields to  $f_{B^0_s}=(10.4^{+1.4}_{-1.3})\%$,
$f_{\mbox{\scriptsize b-baryon}}=(10.4^{+2.2}_{-1.9})\%$ and
$f_{B^0}=f_{B^+} = (39.6^{+1.2}_{-1.4})\%$.
DELPHI has also a preliminary measurement~\cite{delfbs} of the rate 
of charged and neutral weak B hadrons: 
${\cal B}(b \to X_B^0)=(57.8\pm0.5\pm1.0)$\%,
 ${\cal B}(b \to X_B^{\pm})=(42.2\pm0.5\pm1.0)$\%.

\section{Decays}
\subsection{{\boldmath  $b$} decay multiplicity}
There are two new measurements of the mean charged multiplicity
in $b$-hadron decays at the Z, with much smaller systematic uncertainties
than previous measurements. L3 has measured 
$<n_b> = 4.90 \, \pm \, 0.04 \, \pm \, 0.11$~\cite{l3mult};
while DELPHI found $<n_b> = 4.97 \, \pm \, 0.03 \, \pm \, 0.06$~\cite{delmult}.

\subsection{semileptonic branching ratio}
\def \BRb  {${\cal B}(b\rightarrow \ell^-\bar \nu$ X)}
\def \BRc  {${\cal B}(c\rightarrow \ell^+\nu$ X)}
\def \BRbc {${\cal B}(b\rightarrow c\rightarrow \ell^+\nu$ X)}
\def \sBRb  {${\cal B}(b\rightarrow \ell^-$)}
\def \sBRbc {${\cal B}(b\rightarrow c\rightarrow \ell^+$)}

\newcommand {\Prob} {{\cal{P}}(p,p_\perp)}
\newcommand {\xb} {\langle X_b \rangle}
\newcommand {\btol} {b\to\ell}
\newcommand {\btoctol} {b\to c\to\ell}
\newcommand {\brbl}  {${\cal B}(b\rightarrow \ell\mathrm{)}$}
\newcommand {\brbcl} {${\cal B}(b\rightarrow c\rightarrow \ell \mathrm{)}$}
\newcommand {\brcl}  {${\cal B}(c\rightarrow \ell\mathrm{)}$}
\def \QP   {${\rm p}$}
\def \QPT  {${\rm p}_\perp$}
From the experimental point of vue, semileptonic branching ratios
are accessible. They are relatively large and leptons have clean signatures.
Furthermore all detectors have good lepton identification device.
From the theoretical point of vue, 
despite strong interactions are quite important in these decays, they
allow detailed theoretical predictions that can be tested experimentally.
It is why they are among the most extensively studied decays.

At LEP, old analyses were done by performing a multi-parameter fit in the 
(p,p$_{\perp}$) lepton spectrum~\cite{alrbl,delrbl,opalbl}. 
The 4 electroweak heavy quark flavours parameters $R_b$, $R_c$, $A_{FB}^b$,
$A_{FB}^c$ can all be measured simultaneously, with the following parameters:
$\chi$, \brbl, \brbcl, \brcl, $b$, and $c$ fragmentation parameters.
Some collaborations have separate fits for smaller sets of parameters, 
and restrict to a p$_{\perp}$ region to enrich the sample in $b$.
ALEPH has developed new techniques, first presented in 1995\cite{aleps},
 to measure more accurately
\brbl\ and \brbcl\ using information from the silicon vertex detector.
Events are split into 2 hemispheres, and a cut on the
lifetime tag probability\cite{bropap}
is imposed on all hemispheres to prepare a very pure sample of
$Z \to b \bar b$ events. Typically, a purity of 96~\% in $b$ events can be achieved
with an efficiency of 25~\%. Then the opposite hemisphere a tagged hemisphere
is used as an unbiased sample of $b$ decays.
A clear kinematic distinction  allows to disentangle the $b \to \ell$ at high
\QPT\ from $b \to c \to \ell$ at low \QPT.
While single leptons are sufficient to extract ${\cal B}(b \to \ell)$ and
${\cal B}(b \to c \to \ell)$, by performing a fit in the \QPT\ plane,
the opposite-side dilepton sample ,
which is naturally enriched in $b$ decays,
is also used 
to measure at the same time the $b$ fragmentation and the mixing parameter
$\chi$,
taking advantage of the charge correlations.
 The charge correlations allow also to reduce the model dependence.
Although the uncertainty from the semileptonic decay models is
still dominant, it has been reduced nearly by a factor 2.
This new analysis provides some significant improvements in
systematic uncertainties thanks to the use of a very pure sample of $b$ events,
which suppresses the charm and light quark contributions, and to the fact that
 ${\cal B}(b \to \ell)$ is independent of $R_b$ by construction.

DELPHI has performed the same kind of analysis, measuring 
${\cal B}(b \to \ell)$, ${\cal B}(b \to c \to \ell)$ and $\chi$ \cite{delbl}.
And, OPAL has now, for this winter, an analysis of this type~\cite{opalbl}.
They use single muon sample only and  
a neural net to improve the discrimination between $\btol$ and $\btoctol$.
A measurement of ${\cal B}(b \to \ell)$ and ${\cal B}(b \to c \to \ell)$ is
obtained.
All LEP measurements of ${\cal B}(b\rightarrow \ell )$ and 
${\cal B}(b\rightarrow c\rightarrow \ell )$ are summarized 
in table \ref{tabbr}.

The new LEP average value~\cite{LEPEWWG} 
of \brbl\ is 0.1104 $\pm$ 0.0019, while the
$\Upsilon(4s)$ average is 0.1045 $\pm$ 0.0021~\cite{drell}. 
The discrepancy between
these two numbers is 0.0059 $\pm$ 0.0028  corresponding to 2.1 $\sigma$.
Furthermore, beauty baryons are produced at LEP and not at the
$\Upsilon(4s)$ and their semileptonic branching ratio is smaller
(this will be seen later). Consequently  the LEP average is expected to 
be lower
than the $\Upsilon(4s)$ result, contrary to the observed pattern.
If we consider only the last measurements of \brbl, obtained with a
new kind of method and corresponding to a second generation of \brbl\
measurements~\cite{aleps,delbl,opalbl}
which are less model dependent, the Z average becomes
${\cal B}(b\rightarrow \ell )^{\rm Z}$ = 0.1094 $\pm$ 0.0030, to be
compared with $\Upsilon(4s)$~\cite{argbl,clebl} average of 
${\cal B}(b\rightarrow \ell )^{\Upsilon(4s)}$ = 0.1018 $\pm$ 
0.0040~\cite{drell}.
The discrepancy decrease to 1.5 $\sigma$.

\begin{table}[htbp]
\begin{center}
\caption{${\cal B}(b \to \ell)$ and ${\cal B}( b \to
c\to \ell )$ measurements (in \%) at LEP.}
\begin{tabular}{|l|c|c|}
 \hline
        & ${\cal B}(b\to \ell)$                   
& ${\cal B}(b\rightarrow c\rightarrow \ell \mathrm{)}$ \\
\hline
 ALEPH\cite{alrbl} & 11.2  $\pm$ 0.3 $\pm$ 0.4  & 8.8 $\pm$ 0.3 $\pm$ 0.8 \\
 ALEPH\cite{aleps} & 11.0  $\pm$ 0.1 $\pm$ 0.3  & 7.7 $\pm$ 0.2 $\pm$ 0.5 \\
 DELPHI\cite{delrbl} & 11.3  $\pm$ 0.5 $\pm$ 0.7  & 7.9 $\pm$ 0.5 $\pm$ 1.2\\
 DELPHI\cite{delbl} & 10.6  $\pm$ 0.1 $\pm$ 0.4  & 8.3 $\pm$ 0.3 $\pm$ 0.8\\
 L3\cite{l3not2} & 11.4 $\pm$ 0.5 $\pm$ 0.4   &  - \\
 L3\cite{l3bl} & 10.7 $\pm$ 0.1 $\pm$ 0.4   &  - \\
 OPAL\cite{opalpap}& 10.6 $\pm$ 0.6 $\pm$ 0.7   & 8.4 $\pm$ 0.4 
$\pm$ 0.7 \\
 OPAL(n)\cite{opalbl}& 10.9 $\pm$ 0.1 $\pm$ 0.5   & 9.9 $\pm$ 0.3 $\pm$ 1.3 \\ 
\hline
LEP Average  & 11.04 $\pm$ 0.19           & 8.07 $\pm$ 0.34 \\
 \hline
\end{tabular}
\label{tabbr}
\end{center}
\end{table}

Historically, theoretical predictions of the semileptonic $b$ branching
ratio have been significantly larger than the measured values.
Traditionally  ${\cal B}(b \to \ell)^{\rm TH} \ge 12.5 \%$~\cite{bigi}, which
disagrees with the experimental values.
Various aspects of this problem have been scrutinized.
The inclusive semileptonic $b$ branching ratio is defined as:

$${\cal B}(\btol) = \frac{\Gamma_{\rm semi-leptonic}}{\Gamma_{\rm semi-leptonic}
\, + \, \Gamma_{\rm hadronic} \, + \, \Gamma_{\rm leptonic}}$$
with $ \Gamma_{\rm hadronic}\, = \, \Gamma(b \to c \bar u d) \, + \, 
 \Gamma(b \to c \bar c s) \, + \,  \Gamma(b \to {\rm no \, charm}) $
and
$\Gamma(b \to {\rm no \, charm}) \, = \, \Gamma(b \to s(d)\gamma) \, + \, 
 \Gamma(b \to s(d)g) \, + \,  \Gamma(b \to u \bar u d)$.

Solutions of the problem consist to find a way to increase the theoretical
hadronic rate. Theoretical possible solutions are that where
there could be an enhancement of:
\begin{itemize}
\item 
$\Gamma(b \to c \bar u d)$ 
due to non perturbative effects. But, in the same time, these models
predict $\tau_{B^+}/\tau_{B^0} \simeq$ 0.8~\cite{hon}
 which is in desagreement
with the experimental lifetime ratio of 1.07 $\pm$ 0.04 previously
presented in table~\ref{tablif}.
\item 
$\Gamma(b \to c \bar c s)$ due to
large higher order QCD corrections~\cite{bagan1,bagan2,volo}.
In the same time, these models affect the average number of charm quarks 
per $b$ decay, $n_c$, which consequently has to be also measured
experimentally (see later).
\item $b \to$ no open charm, which could be a sizable fraction of
$b \to c \bar c s$ transitions~\cite{dunietz}. 
The hypothesis is that a large component
of low mass $c \bar c$ pairs are seen as light hadrons and not as open charm.
$n_c$ would not be increased by this mechanism.
\item $\Gamma(b \to {\rm no \, charm})$ e.g. large ${\cal B}(b \to s\gamma) $ 
or ${\cal B}(b \to s g) $, from some sources of new physics.
\end{itemize}

{\bf charm counting :} 
Classical charm counting experiments consist in measuring the rates 
of the weakly decaying charm hadrons in selected $b$ events.
The $\Upsilon(4s)$ branching ratio are from CLEO~\cite{cleonc1,cleonc2,cleonc3}
giving $n_c^{\Upsilon(4s)}$ = 1.119 $\pm$ 0.053. LEP measurements are
from ALEPH~\cite{alnc} $n_c$ = 1.230 $\pm$ 0.036 $\pm$ 0.038 $\pm$ 0.053
and from OPAL~\cite{opnc} $n_c$ = 1.061 $\pm$ 0.045 $\pm$ 0.060 $\pm$ 0.037,
where the last error is due to D branching ratio, largely correlated between
the experiments. A main difference between the experiments are assumptions
made about the unmeasured $\Xi_c$ contribution, which is set to 0 in the
case of OPAL, whereas ALEPH estimates it to be 0.063 $\pm$ 0.021.
Accepting this last estimate and including also DELPHI measurements of
$D^0$ and $D^+$ rates~\cite{delnc}, the average result is 
 $n_c^{\rm Z}$ = 1.202 $\pm$ 0.067~\cite{drell,feindt}.

New methods have been developed to estimate $n_c$ which can be written as:
$n_c$ = 1 + ${\cal B}(B\to D \bar D) + {\cal B}(B\to$''hidden'' $c \bar c) - 
{\cal B}(B\to$ no $c)$, where the ''hidden'' $c \bar c$ is the contribution
of bound states (e.g. $J/\psi$). 
DELPHI has determined the fraction of $b$ decays into 0,1 and 2 charmed hadrons
thanks to an analysis of the hemisphere $b$ tagging probability distribution
in terms of Monte Carlo expectations of the three components~\cite{delnc2}.
They have measured:
${\cal B}(b \to 2\,c)  =  0.136 \, \pm \, 0.042 $ and
${\cal B}(b \to 0\,c)  =  0.033 \, \pm \, 0.021 $. 
Substracting the hidden charm contribution of 0.026 $\pm$ 
0.004~\cite{cleonc1,buch} yields a charmless B branching ratio without hidden
charm of ${\cal B}(b \to {\rm no \, charm})$ = 0.007  $\pm$ 0.021, to be
compared with the Standard Model expectation of 0.016 $\pm$ 0.008~\cite{lenz}.
Imposing this SM value they have measured 
$n_c$ = 1.147 $\pm$ 0.041  $\pm$ 0.008.
An upper limit at 95\% CL on new physics in charmless B decays is derived:
 ${\cal B}(b \to {\rm no \, charm})^{\rm New}$ $<$ 0.037.
In another study, correlations of identified charged kaons with inclusively 
reconstructed D mesons were analysed~\cite{delnc3}.
A fit of the transverse momentum
spectra of same and opposite sign K(extra)K(from D) samples has given
${\cal B}(b \to 2\,c)  =  0.170 \, \pm \, 0.035 \, \pm 0.032$ and
${\cal B}(b \to \bar D D_s X)/{\cal B}(b \to 2\,c) = 
0.84 \, \pm \, 0.16 \pm \, 0.09$.

CLEO has selected high momentum leptons, has studied D-lepton angular and charge 
correlations and has looked for wrong sign D~\cite{cleonc4}. 
They have measured the ratio of ``upper vertex'' charm
to ``lower vertex'' charm :
${\cal B}(B\to D X)/{\cal B}(B\to \bar D X)$ =
0.100 $\pm$ 0.026 $\pm$ 0.016. 
From that the number of D's produced at the upper vertex in B decays
${\cal B}(B\to D X)$ = 0.079 $\pm$ 0.022 is derived, as well as
${\cal B}(b \to c \bar c s)$ = 0.219 $\pm$ 0.036 and 
$n_c$ = 1.204 $\pm$ 0.037.

ALEPH has performed an inclusive analysis looking for doubly-charmed B decays,
B $\rightarrow D \bar D (X)$, where D can be either a D$^0$,  D$^+$, D$^{*+}$
or a D$_s$, with both charmed mesons reconstructed~\cite{alnc2}. 
The following branching ratios were measured: 

 $$ \begin{array}{ll}
{\rm {\cal B}(b \rightarrow D_s D^0, D_s D^\pm} (X)) & = \, {\rm 
(13.1^{+2.6}_{-2.2} (stat) ^{+1.8}_{-1.6} (syst) 
       ^{+4.4}_{-2.7}({\cal B}_D))}\% \\

  {\rm {\cal B}(b \rightarrow D^0 \bar D^0, D^0 D^\pm} (X) )& 
= \, {\rm 
 (7.8^{+2.0}_{-1.8}(stat) ^{+1.7}_{-1.5}(syst) 
         ^{+0.5}_{-0.4} ({\cal B}_D))\% } \\
{\rm {\cal B}(b\rightarrow D^{\pm} D^\mp} (X))& < \,  0.9\%\ 
{\rm at\ 90\%\ C.L.} \\
\end{array}$$

\noindent
providing the first evidence for doubly-charmed B decays involving
no ${\rm D_s}$ production.
The sum of the inclusive DD rates is:

\begin{center}
${\rm (20.9^{+3.2}_{-2.8} (stat) ^{+2.5}_{-2.2} (syst) 
       ^{+4.5}_{-2.8}({\cal B}_D)) }\%$
leading to $n_c$ = 1.219 $^{+0.061}_{-0.045}$.
\end{center}

An evidence for associated ${\rm K^0_S}$ and ${\rm K^\pm}$ production
in the decays ${\rm B \rightarrow \bar D D}(X)$ was also found
${\cal B}{\rm (B \rightarrow \bar D^{(*)} D^{(*)} K) = 
 (7.1_{-1.5}^{+2.5}(stat) _{-0.8}^{+0.9}(syst) \pm 0.5({\cal B}_D))
\%. }$ which showed that
${\rm B \rightarrow \bar D^{(*)} D^{(*)} K}$ 
is a large part of ${\rm B \rightarrow \bar D D}(X)$ ($\approx 70\%$).

If previous $n_c$ measurements could show some discrepancies between
results obtained at LEP and at lower energy, the agreement is now better.
Combining all these measurements leads to the world average
$n_c$ = 1.178 $\pm$ 0.021.

{\boldmath  $b \to s \gamma$ :}
The flavour changig neutral current decay $b \to s \gamma$ has been
seen in both exclusive and inclusive channels. The exclusive decay
$B\to K^*\gamma$  has been measured by CLEO~\cite{cleobs} 
${\cal B}(B\to K^*\gamma)$  =  (4.2 $\pm$ 0.8 $\pm$ 0.6) 10$^{-5}$ 
and ALEPH has placed an upper limit on the  $B_s\to \Phi\gamma$ 
penguin decays~\cite{alphi} ${\cal B}(B_s\to \Phi\gamma) <$ 29 10$^{-5}$
at 90\% CL.
CLEO has first observed the inclusive electromagnetic penguin 
decay~\cite{cleobs2} 
${\cal B}(b\to s\gamma)$ = (2.32 $\pm$ 0.57 $\pm$ 0.35) 10$^{-4}$ while
ALEPH has published the first result at LEP. The signal was isolated
in lifetime-tagged $b \bar b$ events by the presence of a hard photon
associated with a system of a high momentum and rapidity hadrons~\cite{albs}.
 ${\cal B}(b\to s\gamma)$ = (3.11 $\pm$ 0.80 $\pm$ 0.72) 10$^{-4}$ was
measured. The average of these two measurements is $(2.54 \pm 0.57) \, 10^{-4}$,
consistent with the Standard Model expectation via penguin 
processes (3.76 $\pm$ 0.30)$10^{-4}$~\cite{thbs}.

{\boldmath  $b \to s g$ :} Large rates of  $b \to s g$~\cite{kagan} 
would show up an extra sources of charged kaons, especially visible at
high momentum in the B rest frame. DELPHI has looked for a high $p_{\perp}$ 
kaon, identified with their RICH or dE/dX in the TPC in $b$ tagged
events~\cite{delnc3}, and derived an upper limit
${\cal B}(b \to s g) < $ 0.05 at 95\% C.L.
In their wrong sign charm paper CLEO~\cite{cleonc4} has derived also
${\cal B}(b \to s g) < $ 0.068 at 90\% C.L.

To summarize, theoretical predictions on \brbl\ can 
accommodate with lower value
as predicted in the framework of 1/$m_Q$ expansions with higher order
perturbative QCD corrections. In these models the rate 
$b \to c \bar c s$ is increased while \brbl\ is decreased.
This is in agreement with the experimental situation.
Other models, which predict new physics for instance, are disfavoured.

{\boldmath  $b$}-{\bf baryon semileptonic branching ratio :}
By determining the ratio $R_{\Lambda \ell} = {\cal B}(\Lambda_b \to
\Lambda \ell^- X)/{\cal B}(\Lambda_b \to \Lambda X)$ = 0.070 $\pm$ 0.012
$\pm$ 0.007 \cite{opalll}, using $\Lambda - \ell $ correlations,
OPAL has a measurement of the $\Lambda_b$ semileptonic branching ratio.
ALEPH, on this side, has a measurement of the $b$-baryon semileptonic
branching ratio, using $p - \ell$ correlations, by determining 
$R_{p \ell} = {\cal B}(b-{\rm baryon} \to p \ell X)/{\cal B}(b-{\rm baryon}
 \to p  X)$
= 0.080 $\pm$ 0.012 $\pm$ 0.014 \cite{alfbar}. 
Both can be assumed to be
very similar to ${\cal B}(b-{\rm baryon} \to \ell)$. They are significantly
lower than the average \brbl. Combining them we get  
${\cal B}(b-{\rm baryon} \to \ell)$ = 0.074 $\pm$ 0.011. This confirms that 
light quarks play a significant role in the decay of $b$-baryons,
as suggested by the short $b$-baryon lifetime measurements presented
earlier. When correlated with this short lifetime, the agreement
between the ratios $ \tau_{{\rm b-baryon}} /  \tau_{B^0_d} = 
 0.78 \pm 0.04$ and ${\cal B}(b-{\rm baryon} \to \ell) / {\cal B}(b \to \ell)
= 0.67 \pm 0.10$ is consistent with the hypothesis of a constant
semileptonic decay width for all $b$-hadrons. 

{\boldmath  $b \to \tau \nu_{\tau} (X)$:}
The study of this channel is interesting because the decay could proceed
by a W or a Higgs boson. Thus a measurement of this decay channel could
be sensitive to new physics, for example supersymmetry where two Higgs
doublets are introduced. The SM prediction for 
${\cal B}(b \to \tau \nu_{\tau} X)$ is (2.30 $\pm$ 0.25)\%~\cite{btausm},
and its measurement limits the ratio tan$\beta$/$m_{H^{\pm}}$,
where  tan$\beta$ is the ratio of the vacuum expectation values
for the Higgs fields and $m_{H^{\pm}}$ is the mass of the
charged Higgs boson.
DELPHI has performed a new measurement~\cite{delbtau}. First $b$-tagging
is used to obtain a sample of $Z \to b\bar b$ events. Then the events are
required to have large missing energy and no electron or muon 
candidates. The result ${\cal B}(b \to \tau \nu_{\tau} X)$ = (2.52 $\pm$
0.23 $\pm$ 0.49)\% is obtained, consistent with previous measurements
from ALEPH: (2.58 $\pm$ 0.19 $\pm$ 0.33)\%~\cite{albtau}, L3:
(1.7 $\pm$ 0.5 $\pm$ 1.1)\%~\cite{l3btau}, and OPAL: (2.58 $\pm$ 0.11
$\pm$ 0.51)\%~\cite{opbtau}. The LEP average is (2.52 $\pm$ 0.26)\%.

The fully leptonic $b \to \tau \nu_{\tau}$ decay is also very interesting.
The expected branching ratio in the SM is 6 $10^{-5}$, however with
large uncertainty. Because of helicity conservation the rates are 
proportional to the square of the lepton mass. The purely leptonic decays
to e and $\mu$ are expected to have the following branching ratios:
 ${\cal B}(b \to e \nu_{e})$ = 5 10$^{-12}$ and
 ${\cal B}(b \to \mu \nu_{\mu})$ = 2 10$^{-7}$.
No signal of $B \to \tau \nu_{\tau}$ decays is observed and upper limits
are given at 90\% CL: ${\cal B}(b \to \tau \nu_{\tau}) < 1.6 \, 10^{-3}$
from ALEPH~\cite{albtau}, 1.1 $10^{-3}$ from DELPHI~\cite{delbtau}
and 5.7  $10^{-4}$ from L3~\cite{l3btau2}. 
From this last limit the best constraint  tan$\beta$/$m_{H^{\pm}}< 0.38$
at 90\% CL is obtainbed.


\subsection{{\boldmath  $|V_{cb}|$} measurements}
There are two main approaches to determine the magnitude of the CKM matrix 
element $|V_{cb}|$, either from inclusive semileptonic B decays or from
exclusive channels such as $B \to D^{(*)} \ell \nu$.

The first one uses the measurement of the inclusive $b$ semileptonic 
branching ratio and has the advantage of great statistical power.
Treating the $b$ quark as a free particle, its semileptonic partial
width is 

$$
\Gamma (b\rightarrow c\ell\nu) = \frac{G_{\rm F}^2\, m_b^5}{192\, \pi^3}
\,\Phi\ |V_{cb}|^2 \equiv \alpha\, |V_{cb}|^2
= \frac{{\cal B}(b\rightarrow c\ell\nu)}{\tau_b} 
$$

where $\Phi$ is a phase space factor. The theoretical dominant uncertainties 
is dominated by the knowledge of the correction due to the
binding of the $b$ quark into a hadron, and the $b$ quark mass dependence.
Recent calculations~\cite{shif}, using HQET in combination with the technique
of  Operator Product Expansion, have rather small
uncertainty and show that they are now under better control.  $|V_{cb}|$
is then given by:
$$
|V_{cb}|~=~0.0419 \sqrt{\frac{{\cal B}(B \rightarrow X_c \ell \bar \nu)}{0.105}} \sqrt{\frac{1.55}{\tau_B}}
(1 \pm 0.015 \pm 0.010 \pm 0.012)
$$
The measured branching ratios need to be corrected for the $b\rightarrow u$
contribution:
$
\frac{{\cal B}(b\rightarrow u\ell\nu)}{{\cal B}(b\rightarrow c\ell\nu)} \simeq
2 \,\left|\frac{V_{ub}}{V_{cb}}\right|^2 = (1.5\pm 1.0)\%\  .
$
The value of  $|V_{cb}|$~\cite{drell} is 
(38.7 $\pm$ 2.1) $10^{-3}$ at the $\Upsilon(4s)$ and 
(40.6 $\pm$ 2.1) $10^{-3}$ at the Z.

The second technique for measuring  $|V_{cb}|$ is based on the study of
exclusive channels such as $B \to D^{(*)} \ell \nu$. It has less
theoretical limitations. The rate of this process is governed by  $|V_{cb}|$
and Heavy Quark Symmetry provides model independent relations between
the relevant weak decay form factors in the heavy quark limit.
The corrections from the heavy quark symmetry breaking are calculable
in the framework of HQET. The differential decay rate of 
 $B\rightarrow D^{(*)}\ell\nu$, with respect to the boost $w$ 
($w = (m_B^2 + m_D^2 - q^2)/(2\, m_B\, m_D)$) of the 
$D^{(*)}$ in the $B$ rest frame, is given by
$
{\rm d}\Gamma(B\rightarrow D^{(*)}\ell\nu)/
{\rm d}w = G(w)\,|V_{cb}|^2 {\cal F}^2(w)
$
where $G(w)$ is a known phase space function and ${\cal F}(w)$ is a universal
hadronic form factor.   ${\cal F}(w)$ parametrizes the effects of the strong 
interaction on the decay, with an unknown shape.
The product $|V_{cb}| {\cal F}(w)$ is then extrapolated to $w=1$,
which corresponds to the maximal value of $q^2$, by the expansion
$
{\cal F}(w) = {\cal F}(1)\,\left[ 1 - \rho^2(w-1) + c(w-1)^2 + ... \right]\ .
$
The intercept and slope are strongly correlated, and this needs to be
accounted for when averaging results of different experiments.
Furthermore, the expected value of ${\cal F}(1)$ is not exactly unity
due to correction for the finite heavy quark mass: 
${\cal F}(1)_{D^* \ell \nu} = \, 0.91 \, \pm \, 0.03 $ for
 $B\rightarrow D^*\ell\nu$ decays~\cite{shif1}, while 
${\cal F}(1)_{D \ell \nu} = \, 0.98 \, \pm \, 0.07$ for 
 $B\rightarrow D\ell\nu$ decays~\cite{neub1}.
The decay $B\rightarrow D^*\ell\nu$ is favoured for the measurement, as 
the $1/m_Q$ correction is predicted to vanish, and experimentally
it has a large branching ratio and clean signal. There is a new preliminary
analysis from DELPHI, using a new parametrization~\cite{neub} giving a
more precise expansion versus the axial form factor,
${\cal A}(w)  = {\cal A}(1)\,\left[ 1 + \rho^2_{\cal A}{\cal G}(w) 
+ ... \right]\ .$ (${\cal A}(1)  \equiv {\cal F}(1)$).
They got
$
{\cal A}(1) V_{cb}  = (37.7 \, \pm  \, 1.7 (stat) \, \pm \, 
1.7 (syst)) \, 10^{-3}$, 
$\rho_{\cal A}^2  = 1.36 \, \pm \, 0.17 \, \pm \, 0.14$, 
${\cal B}(B^0 \to \ell \nu D^*)  =  (5.18 \, \pm \, 0.16 \, \pm \, 0.49) \% $
and $|V_{cb}|$ = (41.4 $\pm$ 3.0) 10$^{-3}$.
We can hope that in the future existing data will be re-analysed
using this new parametrization.
The previous published measurements are presented in table~\ref{tabvcb}
and the average of the $V_{cb}$ measurements from this channel,
 $B\rightarrow D^*\ell\nu$, is  
$|V_{cb}|$ = (38.7 $\pm$ 3.1) 10$^{-3}$~\cite{gibbons}.
\begin{table}[htbp]
\begin{center}
\caption{Experimental values of  $|V_{cb}| {\cal F}(1)$.}
\begin{tabular}{|l|c|c|}
 \hline
& $B\rightarrow D^*\ell\nu$ & $B\rightarrow D \ell\nu$ \\
\hline
 ALEPH\cite{alvcb} & (32.1 $\pm$ 1.8 $\pm$ 1.9) $10^{-3}$ &
 (28.2 $\pm$ 6.8 $\pm$ 6.5) $10^{-3}$ \\
 ARGUS\cite{argvcb} & (39.2 $\pm$ 3.9 $\pm$ 2.8) $10^{-3}$ &\\
 CLEO\cite{clvcb} & (35.2 $\pm$ 1.9 $\pm$ 1.8) $10^{-3}$ &
 (34.2 $\pm$ 4.4 $\pm$ 4.9) $10^{-3}$ \\
 DELPHI\cite{delvcb} &  (36.9 $\pm$ 2.1 $\pm$ 2.2) $10^{-3}$ & \\
 OPAL\cite{opvcb}&  (32.6 $\pm$ 1.7 $\pm$ 2.2) $10^{-3}$ & \\
 \hline
\end{tabular}
\label{tabvcb}
\end{center}
\end{table}

The decay $B\rightarrow D \ell\nu$ can also be used to measure 
 $|V_{cb}|$ in a similar manner. Here there are fewer experimental
results, as it is a more challenging mode, and only ALEPH~\cite{alvcb} 
and CLEO~\cite{clvcb}
have used this channel (see table~\ref{tabvcb}). Combining their
results gives $|V_{cb}|$ = (39.4 $\pm$ 5.0) 10$^{-3}$~\cite{gibbons}.
The experimental values for the intercept and the slope are combined with
careful attention to the correlated errors.

The excellent agreement between a wide variety of methods for
extracting $V_{cb}$ is encouraging. The world average of all
$V_{cb}$ measurements leads to $|V_{cb}|$ = (39.5 $\pm$ 1.7) 10$^{-3}$.
Improvements from theoretical and experimental sides are promising.

\subsection{{\boldmath  $|V_{ub}|$} measurements}
 $|V_{ub}|$ can be extracted in a same way as  $|V_{cb}|$ by looking
to $b \to u$ transitions instead of $b \to c$ transitions.
The inclusive and exclusive approaches can also be used.
Due to the fact that a $b$ decay is dominated by  the process where
the $b$ quark turns into a $c$ quark, the experimental determinations of
 $|V_{ub}|$ are very much difficult than  those of $|V_{cb}|$. It is also
more difficult from the theoretical point of vue because the $u$ quark
in the final state is no longer heavy.

The first observations for  $b \to u$ transitions were done at the 
$\Upsilon(4s)$ were $B$ mesons are produced at rest~\cite{vub}.
These analyses have looked at the endpoint of the single lepton
spectrum for leptons from $B$ decay that are kinematically incompatible
with coming from the decay of a $B$ meson to charm meson. From the lepton
excess in this corner of phase space, theoretical models are used to
extrapolate the full lepton spectrum and   $|V_{ub}|$ = (3.1 $\pm$ 
0.8)$10^{-3}$ is extracted by CLEO~\cite{cleovub}. Model uncertainties
dominate the error.

ALEPH has published the first evidence for semileptonic $b \to u$ transitions
in $b$ hadrons produced at LEP~\cite{alvub}.
ALEPH has inclusively reconstructed the hadronic system accompagning the
lepton in the semileptonic B decays and has built a set of kinematic
variables to discriminate between $X_u \ell \nu$ and  $X_c \ell \nu$
transitions by taking advantage of the different shape properties of these
final states. A neural network was used to extract the inclusive
${\cal B}(b \to X_u \ell \nu)$ branching ratio.
They have measured ${\cal B}(b \to X_u \ell \nu)$ = 
(1.73 $\pm$ 0.55 $\pm$ 0.55) 10$^{-3}$. 
An advantage of this analysis is that it integrates over the entire lepton
and hadron spectrum for these decays, potentially reducing the model
dependence of the result. A disadvantage is that it has to manage
with the large background from $b \to c$ semileptonic decays which needs
to be well understood.

The same kind of calculations than for  $|V_{cb}|$ have been done to
extract  $|V_{ub}|$ from ${\cal B}(b \to X_u \ell \nu)$~\cite{shif}
$$
|V_{ub}|~=~0.0465 \sqrt{\frac{{\cal B}(B \rightarrow X_u \ell \bar \nu)}{0.002}} \sqrt{\frac{1.55}{\tau_B}}
(1 \pm 0.025_{\rm pert} \pm 0.03_{m_b})
$$
The resulting value of  $|V_{ub}|$ is (4.16 $\pm$ 1.02)  10$^{-3}$.

 $|V_{ub}|$ has also been extracted from the exclusive decays
$B \to \pi \ell \nu$ and $B \to \rho \ell \nu$ at CLEO~\cite{cleovub2}.
The value is $|V_{ub}|$ = (3.3 $\pm$ 0.2 $\pm$ 0.4 $\pm$ 0.7) 10$^{-3}$,
where the errors are statistical, experimental systematic, and theoretical
model dependence respectively.

All these  $|V_{ub}|$ extractions are model dependent, however in
different ways. The consistency of the results is thus comforting.
The world average is then  $|V_{ub}|$ = (3.4 $\pm$ 0.5 ) 10$^{-3}$

\section{Summary and prospects}
$b$ physics is a place of intensive work. The $e^+ e^-$ colliders
and Tevatron have provided a large amount of data on the production and
decay of beauty particles. In recent years the experiments at LEP, SLC
and CESR have turned the studies on the $b$ quark into precision physics.
Numerous interesting new results have been provided and
a lot of measurements have been performed.
Among these results:
\begin{itemize}
\item In the electroweak sector, $R_b$ has been measured, 
$R_b^0$ = 0.2173 $\pm$ 0.0009, to an accuracy of  $\sim 0.4 \%$,
the forward backward charge asymmetry 
 $A^{0,b}_{FB}$ = 0.0998 $\pm$ 0.0022 leads to a measurement of
$\sin^2 \theta_{\rm eff}$= 0.23213 $\pm$ 0.00039 with an accuracy of
$\sim 0.17 \% $. One of the most precise determination of this quantity.
\item In bottom spectroscopy, the $B_c$ meson has been discovered by
CDF at Tevatron.
\item The $B^+$, $B^0$, $B^0_s$ and $b$-baryon lifetimes, and their ratio
with the  $B^0_d$ lifetime, have been measured with a good precision
($\sim 5 \%$). The  hierarchy among $b$ hadron lifetimes agrees with 
theoretical predictions. But for the $b$ baryon lifetime, predictions are 
higher than measurements by $\sim 3\sigma$. The semileptonic $b$ baryon
branching ratio confirms this puzzle ...
\item The \bd-\bdbar\ oscillation frequency, $\Delta m_d  =  0.466 \, \pm \, 
0.018 \, ps^{-1}$, is also measured with similar accuracy ($\sim 4\%$);
but the hadronic uncertainty limits the extracted value of
$|V_{\rm td}|  =   (8.8 \pm 0.2_{\Delta m_{\rm d}}
{\small
\mp 0.2_{m_{\rm t}} {\mp^{1.4}_{1.8}}_{F\sqrt{B}})} \times 10^{-3} $
to an accuracy of $\sim 20 \%$.

The \bs-\bsbar\ oscillation is still not seen and a limit $\Delta m_s  >  10.2 
\, ps^{-1}$ at 95\% CL has been set. Nevertheless non negligible
constraints on the CKM matrix with
$
 \frac{\displaystyle |V_{\rm ts}|}{\displaystyle |V_{\rm td}|}   >   3.8$ 
at 95\% CL is provided.
\item $b$ semileptonic decays have been scrutinized intensively.
If the situation between ${\cal B}(b \to \ell)$ measurements at $\Upsilon(4s)$
and Z energies is still to be clarified, the agreement being of the order of
2$\sigma$, these low  ${\cal B}(b \to \ell)$ values agree with theoretical
expectations which favour high values of the number of charm quark per
$b$ decay, as experimentally measured. A significant contribution of
decays in $b \to c \bar c s$ transition, without $D_s$ is observed.
No hints for new physics is found in $b$ decays.
\item The CKM matrix element  $|V_{cb}|$ = (39.5 $\pm$ 1.7) 10$^{-3}$
has been measured to an accuracy of $\sim 4\%$. Values of $|V_{cb}|$ 
show good agreement in results obtained from inclusive and exclusive studies.
$|V_{ub}|$ = (3.4$\pm$ 0.5) 10$^{-3}$ has also been measured but to an 
accuracy of $\sim 15\%$ as it is a lot more challenging. Limitations
are due to models.
\end{itemize}
The accuracy of present data provides important tests of the SM and guides
the developements of the understanding of the rich phenomenology of
weak $b$ decays.

No more substantial running at the Z pole is foreseen at LEP and the final
analyses with optimised algorithms, or new analysis ideas, are being
prepared. LEP has shown its capability for studying $b$ physics.
The large amount of data, highly efficient detectors, and good particle
identification have allowed a wide range of $b$ physics to be explored. The 
Z pole is very competitive to study $b$ physics.
SLC is still running and can beneficts for the polarization and excellent
resolution of their vertex detector to provide interesting results.
CESR will continue to improve its luminosity and CLEO has upgraded the
detector and will continue to play a significant role in $B$ physics.
The experiments at the $e^+ e^-$ colliders have also prepared important
engineering data and developed analysis techniques that will be further
exploited in the continuation of beauty physics at the next generation of
$b$ facilities.

HERA-B, at DESY, will start to collect data in 1998.
Next year, in 1999, BABAR and BELLE will start to take data in the new
$b$-factories. CDF and D0 get upgraded and will take data at much 
higher luminosity at the Tevatron, with perhaps B-TEV in a few years.
Then the LHC experiments will enter the scene in 2005, in particular
LHCb, a specific detector to study $b$ physics at $pp$ collider,
which will be able to do many analyses with large precision.

In conclusion there  is a bright future for $b$ physics in the next decade,
with new dedicated or upgraded colliders and detectors, and 
with some challenges. The first observation of \bs-\bsbar\ oscillations and the
measurement of $\Delta m_s $, the study of rare decays and of
CP violation ...

\section*{Acknowledgements}
I would like to thank the ALEPH, CDF, CLEO, DELPHI, D0, L3, OPAL and SLD
collaborations, and their representatives who provided results for this review;
and the LEP Electroweak Heavy Flavour, B-Lifetime and B-oscillation working
groups. I am also grateful to all the people who help me in preparing this
review, and in particular Z. Ajaltouni, C. Ferdi, S. Monteil, V. Morenas,
P. Rosnet, O. Schneider, ...
I would also like to thank the organisers of this conference.


\end{document}